\documentclass{IOS-Book-Article}

\usepackage{mathptmx}
\usepackage{soul}\setuldepth{article}
 
\usepackage[utf8]{inputenc}
\usepackage{graphicx}

\usepackage{tabulary}

\usepackage{comment}

\usepackage{adjustbox}

\usepackage{multirow,booktabs,setspace,caption}
\usepackage{tikz}
 
\def\hb{\hbox to 10.7 cm{}}

\graphicspath{{images/}}

\begin{document}

\pagestyle{headings}
\def\thepage{}

\begin{frontmatter}              

\title{Anomalous Sound Detection with Machine Learning: A Systematic Review\\}

\markboth{}{January 2021\hb}

\author[A]{\fnms{Eduardo Carvalho} \snm{Nunes}%
\thanks{Corresponding Author: Eduardo Carvalho Nunes, ALGORITMI Centre, Department of Information Systems, University of Minho, 4804-533 Guimarães, Portugal; E-mail:
id9303@alunos.uminho.pt}},


\runningauthor{B.P. Manager et al.}
\address[A]{ALGORITMI Centre, Department of Information Systems, University of Minho, Guimarães, Portugal}

\begin{abstract}
Anomalous sound detection (ASD) is the task of identifying whether the sound emitted from an object is normal or anomalous. In some cases, early detection of this anomaly can prevent several problems. This article presents a Systematic Review (SR) about studies related to Anamolous Sound Detection using Machine Learning (ML) techniques. This SR was conducted through a selection of 31 (accepted studies) studies published in journals and conferences between 2010 and 2020. The state of the art was addressed, collecting data sets,methods for extracting features in audio, ML models, and evaluation methods used for ASD. The results showed that the ToyADMOS, MIMII, and Mivia datasets, the Mel-frequency cepstral coefficients (MFCC) method for extracting features, the Autoencoder (AE) and Convolutional Neural Network (CNN) models of ML, the AUC and F1-score evaluation methods were most cited.

\end{abstract}

\end{frontmatter}
\markboth{January 2021\hb}{January 2021\hb}

\section{Introduction}
Anomaly Sound Detection (ASD) has received a lot of attention from the scientific machine learning community in recent years \cite{r1}. An anomaly in sound can indicate an error or defect, detecting the anomaly earlier can avoid a series of problems such as industrial equipment predictive maintenance and audio surveillance of roads \cite{r2, r3}.

Anomaly detection techniques can be categorized as supervised anomaly detection, semi-supervised anomaly detection, and unsupervised anomaly detection. Supervised anomaly detection requires the entire dataset to be labeled "normal" or "abnormal" and this technique is basically a type of binary classification task. Semi-supervised anomaly detection requires only data considered "normal" to be labeled, in this technique, the model will learn what "normal" data are like. Unsupervised anomaly detection involves unlabeled data. In this technique, the model will learn which data is "normal" and "abnormal" \cite{r4}.

This paper presents a systematic review (SR) aiming to verify the state of the art in audio anomaly detection using machine learning techniques. Additionally, it was analyzed which datasets, methods for extracting features in audio, ML models, and evaluation methods most used in the accepted primary studies.Thus, this survey can enable a general analysis of the scope of the work.

In addition to this introductory section, the paper is organized as follows. The Research Methodology section presents the concept of SR, the defined and used protocol, and the process of conducting the review. The "Results and discussion" section presents and discusses the results.


\section{Research Methodology}
\begin{figure}[h!]
	\centering
	\includegraphics[width=1\linewidth]{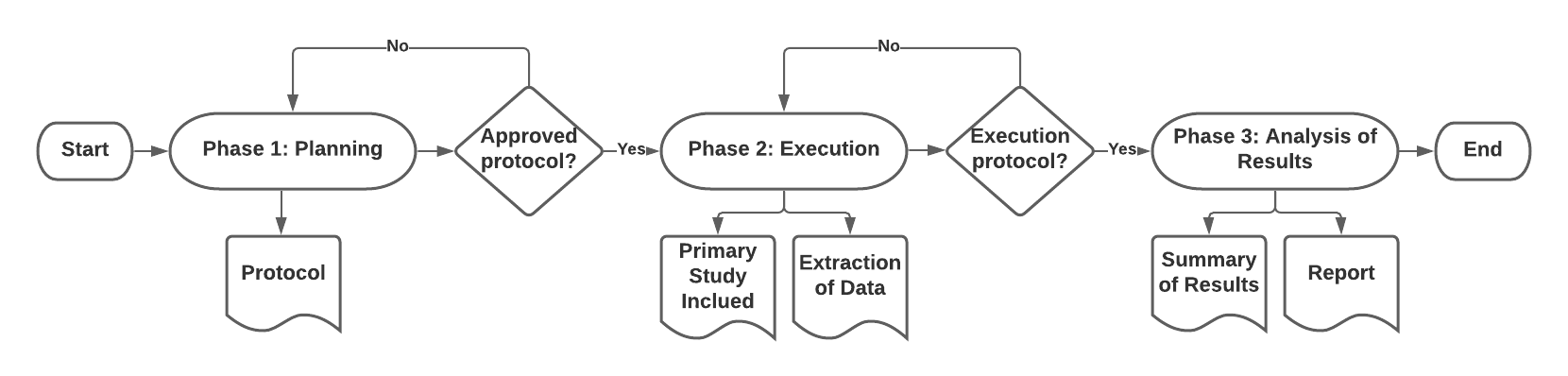}
	\caption{RS phases adapted \cite{r7}.}
	\label{fig:arquitetua proposta}
\end{figure}

Unlike traditional literature reviews, the SR is a rigorous and reliable research method that aims to select relevant research, collect and analyze data, and allow evaluation \cite{r5}. According to Kitchanhan's suggestion, this paper was developed considering the 3 phases: planning, execution and analysis of results (Figure \ref{fig:arquitetua proposta}) \cite{r6}. In the planning phase, a protocol is defined specifying research questions, keywords, inclusion and exclusion criteria for primary studies and other topics of interest. In the execution phase, the bibliographic research is conducted following the defined protocol, it is in this phase that the inclusion and exclusion of primary studies is done. And finally, the results analysis phase, the extraction of the data is done and the results are compared.

\subsection{Planning}
First, a detailed protocol was designed to describe the process and method to be applied in this SR (Table \ref{tab:protocol}).
This protocol contains: objective, main question, keywords and synonyms, study language, sources search methods, study selection criteria, source list, quality form fields and data extraction form fields.

\begin{table}[htpb]
\caption{Defined Protocol for this SR.}
\begin{tabular}{p{0.2\textwidth}p{0.7\textwidth}}
\toprule
\midrule
    {\bf Objective} & This Systematic Literature Review Protocol (SLRP) presents the methodological structure for the implementation of the literature review stage on audio anomaly detection with machine learning techniques. \\
    
    \hline
    
    {\bf Main Question} & What machine learning techniques for audio anomaly detection? \\
    
    \hline
    
    {\bf Keywords and Synonyms } & Anomalous Sound Detection; Anomaly Detection; Detecting Anomalous Audio; Detection of Anomalous Sounds; Machine Learning; Self-Supervised Anomaly Detection; Semi-Supervised Anomaly Detection; Unsupervised Anomaly Detection; Audio; Sound.\\
    
    \hline
    
    {\bf Study Language} & English. \\
    
    \hline
    
    {\bf Sources Search Methods} & The sources should be available via the web, preferably in scientific databases in the area (computer science, computing, electronics). In addition to traditional databases, some can be included according to the results found. Primary studies in other media can also be selected, as long as they meet the requirements of the SR.

    This process will be carried out by means of searches composed of keywords. Primary studies will be found from searches carried out on search portals for articles, theses, dissertations, and journals.
    
    During the information retrieval procedure, the strings found preferably in Titles, Abstracts, and Keywords of each database will be considered.
    
    After checking the relevance of the work, it will be selected for reading (full text). Primary studies will then be accepted or rejected. There will be (I) Inclusion and (E) Exclusion criteria for each primary study analyzed. \\
    
    \hline
    
    {\bf Study Selection Criteria} & 
      
    \textbf{Inclusion}: Anomaly detection in audio; uses machine learning technique; primary study is written in english.
    
     \textbf{Exclusion}: Not detect anomaly in audio; not uses a machine learning technique; it is not written in english; full paper not found; not present the abstract; have a publication year outside the specified deadline (i.e., earlier than 2010). \\
     
     \hline
     
    {\bf Sources} & In addition to the below sources, a search was made for papers in the research community on Detection and Classification of Acoustic Scenes and Events (DCASE).
    
    {\bf ACM }: http://portal.acm.org
    
    {\bf IEEE }: https://ieeexplore.ieee.org/Xplore/home.jsp/
    
    {\bf SCOPUS }: http://www.scopus.com
    
    {\bf DCASE2020}: http://dcase.community/ \\
    
    \hline
    
    {\bf Quality Form Fields} & 
    
    Coherence and cohesion textual;
    
    Uses machine learning technique in an objective way;
    
    Machine learning techniques are cited;    \\
    
    \hline
    
    {\bf Extraction Form Fields} & 
    Which Machine learning category? 
    
    Anomaly detection category? 
    
    Which dataset used?
    
    Which programming language is used?
    
    Which libraries or structure used?    \\
    
    \hline
    
\bottomrule
\label{tab:protocol}
\end{tabular}
\end{table}

\subsection{Selection}
The searches were carried out between November and December 2020. Only recent studies (published since 2010) were considered for assessing the state of the art. The primary studies returned from the electronic database were identified through the search string:

\emph{("anomalous sound detection" OR "detecting anomalous audio" OR "detection of anomalous sound" OR "anomaly detection") AND ("machine learning" OR "supervised anomaly detection" OR "semi-supervised anomaly detection" OR "unsupervised anomaly detection") AND ("sound" OR "audio")}

\begin{figure}[h!]
	\centering
	\includegraphics[width=0.5\linewidth]{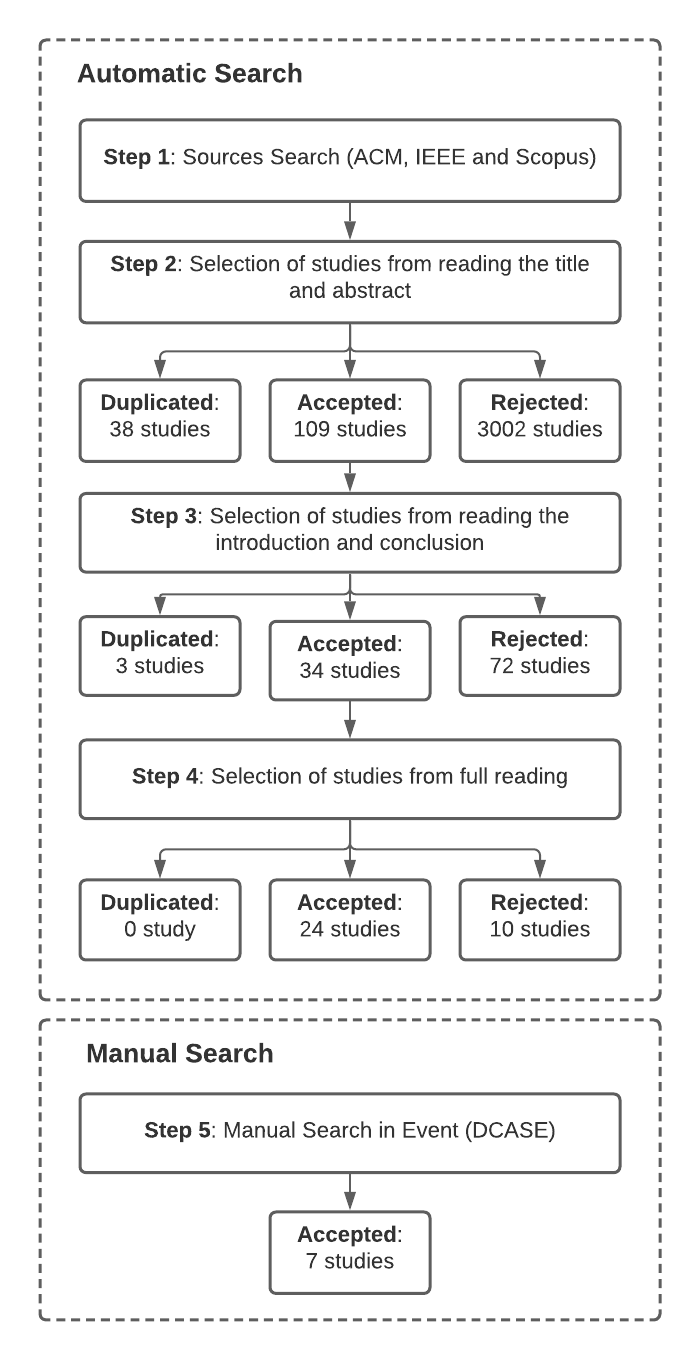}
	\caption{Selection Process of Primary Studies.} 
	\label{fig:primarystudy}
\end{figure}

The selection process for primary studies is illustrated in Figure \ref{fig:primarystudy}. In the first step, 3150 primary studies were identified. In the second step, the titles and abstracts were read and the inclusion and exclusion criteria were applied. In this step, 109 studies were accepted, 3002 studies were rejected and 38 studies were duplicated. In the third step, the introduction and conclusion were read and the inclusion and exclusion criteria were also used. In this step, 34 studies were accepted, 72 studies were rejected and 3 duplicated studies. In the fourth step, 25 studies fully were read and 25 studies were accepted. After completing the selection of studies, it was noted that DCASE was widely cited. With that, a manual search was made and 7 more studies were accepted.

\subsection{Analysis of Results}
This phase consists of a review and extraction of information. The Table \ref{tab:allpapers} shows the number of primary studies collected in each indexed database. It is important to note that Scopus covers some results from the ACM and IEEE. For each primary study, a summary was written with the main study topics.

\begin{table}[htbp]	
	\centering
	\caption{\small{ Number of studies obtained in the indexed databases.}}
	{\small
		\begin{tabulary}{\columnwidth}{lccc}	
			\hline 	
			{\bf Source} & {\bf Nº of Studies} & {\bf Accepted - Selection Phase} & {\bf Accepted - Extraction Phase}   \\ 
			\hline
			ACM  & 376 & 10 (2.65\%) & 2 (20\%)\\
			
			IEEE & 1100 & 25 (2.27\%) & 3 (12\%)\\
			
			SCOPUS & 1674 & 74 (4.42\%) & 19 (25.67\%) \\
			
			DCASE & 49 & 7 (14.28\%) & 7 (100\%)\\
			
			Total & 3199 & 116 (3.62\%) & 31 (26.72\%) \\
			
			\hline
		\end{tabulary} 
	}	
	\label{tab:allpapers}
\end{table}

In the selection phase, 116 studies were selected for the next phase. After completing the selection phase, 31 studies were selected for the extraction phase. It is important to note that the selection of studies related to DCASE, was done manually, that the studies of the selection phase were all accepted for the extraction phase. The main topics of interest: Machine Learning Technique, Anomaly Type, Dataset, Audio Feature Extraction Method, Anomaly Detection Model, and Machine Learning Model Evaluation Method. The main results of SR are described in the results section.

\subsection{StArt Tool}
The StArt tool (State of the Art through Systematic Review) was used to support the SR process \cite{r20}. This tool was created by LaPES-Software Engineering Research Lab (Federal University of São Carlos) and was developed with the purpose of to automate the phases of SR. The tool offers full support for SR and is divided into: Planning, Execution, and Summary.

\section{Results and Discussions}
\subsection{Journals and Proceedings}
 All primary studies were retrieved from scientific journals and conference proceedings. The Table \ref{tab:journals} lists primary studies published in journals. In general, the journals affiliated to IEEE obtained more studies with 5 (50\%) primary studies, two of which have the best impact factor. The Table \ref{tab:procee} lists the primary studies published in Proceedings. As shown in the table, the DCASE 2020 contain more studies with 7 (33\%) primary studies. It is important to highlight that in the event there was a competition (task 2) that is totally related to the detection of anomalies in audio. Another important observation is that the proceedings affiliated with IEEE, which had 8 primary studies (38\%) accepted in this SR.

\begin{table}[htbp]	
	\centering
	\caption{\small{Journals that provided the included primary studies.}}
	{\small
		\begin{tabulary}{\columnwidth}{lcc}	
			\hline 	
			{\bf Journal} & {\bf Impact Factor} & {\bf Nº of Studies}   \\ 
			\hline
			IEEE Transactions on Intelligent Transportation Systems  & 6.319 & 1 \\
			(ISSN 1558-0016) & & \\
			
			IEEE Transactions on Information Forensics and Security & 6.013 & 1 \\
			(ISSN 1556-6021) & & \\
			
			Engineering Applications of Artificial Intelligence & 4.201 & 2 \\
			(ISSN: 0952-1976) & & \\
			
			IEEE Access (ISSN 2169-3536) & 3.745 & 1 \\
			
			IEEE/ACM Transactions on Audio, Speech,  & 3.398 & 2 \\
			and Language Processing (ISSN 2329-9304) & & \\
			
			Electronics (ISSN 2079-9292) & 2.412 & 1 \\
			
			Expert Systems (ISSN 1468-0394) & 1.546 & 1 \\
			
			\hline
		\end{tabulary} 
	}	
	\label{tab:journals}
\end{table}

\begin{table}[htbp]	
	\centering
	\caption{\small{Proceedings that provided the included primary studies.}}
	{\small
		\begin{tabulary}{\columnwidth}{lc}	
			\hline 	
			{\bf Proceeding} & {\bf Nº of Studies}   \\ 
			\hline
		    Proceedings of the Fifth Workshop on Detection and Classification of  &  7\\
		    Acoustic Scenes and Events(DCASE 2020) & \\
		    
			IEEE International Conference on Acoustics, Speech, and Signal  &  4 \\
			Processing (ICASSP) & \\
			
			IEEE Workshop on Evolving and Adaptive Intelligent Systems (EAIS) & 1 \\
			
			IEEE International Conference on Machine Learning and & 1 \\
			 Applications (ICMLA) & \\
			
			IEEE Workshop on Machine Learning for Signal Processing & 1 \\
			
			IIEEE International Conference on Advanced Trends in  & 1 \\
			Information Theory (ATIT) & \\
			
			International Conference on Industrial Engineering and Applications (ICIEA)  & 1 \\
			 
			International Conference on Soft Computing Models in Industrial and  & 1 \\
			Environmental Applications(SOCO) & \\
			
			European Signal Processing Conference (EUSIPCO) & 1 \\
			
			International Conference on Intelligent Networking and  & 1 \\
			Collaborative Systems (INCOS) & \\
			
			Proceedings of the International Conference on Pattern Recognition and  & 1 \\
			Artificial Intelligence & \\
			
			International Conference on Soft Computing and Machine Intelligence  & 1 \\
			 (ISCMI) & \\
			
			\hline
			
		\end{tabulary} 
	}	

	\label{tab:procee}
\end{table}

 The studies were published by 133 different researchers. Table \ref{tab:allauthor1} (Appendix C) shows a summary of the researchers responsible for two or more studies. In this table we can see that the highlights are laboratories from Japan. Figure \ref{fig:allautors} (Appendix B) shows the total number (per country) of researchers with published studies. Highlights Japan with 38 (28\%) researchers and Italy with 29 (22\%) researchers.
 
 \begin{figure}[!ht]
	\centering
	\includegraphics[width=0.8\linewidth]{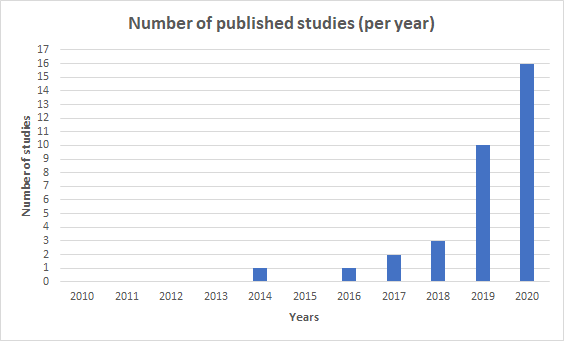}
	\caption{Number of primary studies published per year.}
	\label{fig:studiespubperyear}
\end{figure}

 Figure \ref{fig:studiespubperyear} shows the evolution of the research areas in relation to the number of published studies. According to this SR, the theme of this study had its first study published only in 2014. In the years 2017 and 2018, there were few published studies. However, in 2019 and 2020 many more studies began to emerge.

\subsection{Primary studies accepted in the SR}
Tables \ref{tab:papersRS1}, \ref{tab:papersRS2}, and \ref{tab:papersRS3} shows a synthesis of the 31 primary studies analyzed including: dataset, audio features, ML model and evaluation method for each one. A global analysis is presented in this session taking into account the most relevant topics.

\begin{table}[htpb]
\caption{Anomaly Detection in audio presented in the included studies of the SR (Automatic Search).}
\begin{tabular}{p{0.05\textwidth}p{0.05\textwidth}p{0.15\textwidth}p{0.2\textwidth}p{0.2\textwidth}p{0.15\textwidth}}
\toprule
    {\bf Study} & {\bf Year} & {\bf Dataset} & {\bf Audio Features} & {\bf ML Model} & {\bf Evaluation Method} \\
\midrule
     \cite{rs1} & 2020 & Mivia Dataset \cite{r8} & STFT, MFCC, Mel-Scale & DenseNet-121, MobileNetV2, ResNet-50  & RR, MDR, ER, FPR, Accuracy \\
			
	\cite{rs2} & 2020 &  ToyADMOS \cite{r9},MIMII \cite{r10} & Mel-Filterbank & SPIDERnet, AE, Naive MSE, PROTOnet & AUC, ROC, TPR, FPR, F-measure \\
	
	\cite{rs3} & 2020 & Mivia Dataset \cite{r8} & Audio Power, Audio harmonicity, Total loudness in Bark scale, Autocorrelation coefficient, ZCR, Log-attack time, Temporal centroid, Audio spectrum roll-off, Audio spectrum spread, Audio spectrum centroid, MFCC, Audio spectral flatness & one-class SVM, and DNN & Accuracy, F1-score, Precision\\

	\cite{rs4} & 2020 & Own Dataset & MFCC, and Mel filterbank energies & LSTM & Accuracy, and F1-score \\
	
	\cite{rs5} & 2020 & Own Dataset, Freesound \cite{r10} & MFCC, DWT, ZCR, SR, and GFCC & SVM, Random Forest, CNN, KNN Gradient Boosting, & Precision, Recall, F1-score, Accuracy,  p-value\\
	
	\cite{rs6} & 2020 & Own Dataset & Mel-spectrogram & Conv-LSTM AE, and CAE  & ROC-AUC, F1-score \\
	
	\cite{rs7} & 2020 & Own Dataset & Mel-spectrogram & CAE, and One-Class SVM & ROC-AUC, F1-score \\

	\cite{rs8} & 2020 & Mivia Dataset \cite{r8} & Gammatonegram images & AReN (CNN) & Accuracy, RR, MDR, ER, FPR \\
	
	\cite{rs9} & 2019 & Own Dataset & Mel-spectrogram & Deep AE &  ROC-AUC curve\\
	
	\cite{rs10} & 2019 & Toy Car Running  Dataset \cite{r11} &  Time-series of acoustic & AE & AUC\\

	\cite{rs11} & 2019 & UrbanSound8K \cite{r12}, TUT Dataset \cite{r13} & LPC, MFCC, and GFCC & Agglomerative Clustering, BIRCH & Precision, Recall, F1-score, TP, FP, FN\\
	
	\cite{rs12} & 2019 &  TUT Dataset \cite{r13} & Raw audio & WaveNet, and CAE & ROC-AUC curve\\

\bottomrule
\label{tab:papersRS1}
\end{tabular}
\end{table}    

\begin{table}[htpb]
\caption{Continued. Anomaly Detection in audio presented in the included studies of the SR (Automatic Search).}
\begin{tabular}{p{0.05\textwidth}p{0.05\textwidth}p{0.15\textwidth}p{0.2\textwidth}p{0.2\textwidth}p{0.15\textwidth}}
\toprule
    {\bf Study} & {\bf Year} & {\bf Dataset} & {\bf Audio Features} & {\bf ML Model} & {\bf Evaluation Method} \\ 
\midrule
    \cite{rs22} & 2019 & DCASE2018 Task 1 \cite{r17}, DCASE2018 Task 2 \cite{r18} & FFT, and Log mel spectrogram & CNN & F1-score, AUC, mAP, AP, ER\\ 
    
    \cite{rs24} & 2019 &  TUT Dataset \cite{r13}, NAB Data Corpus \cite{r19} & Raw data &One-Class SVM, and LSTM-AE & Accuracy \\

    \cite{rs13} & 2019 &  Own Dataset &  Log mel energy & Deep AE & AUC \\
    
    \cite{rs14} & 2019 & Own Dataset, Effects Library \cite{r14}  &  Log mel spectrum, MFCC, General Sound, i-vector & WaveNet, AE, BLSTM-AE, AR-LSTM & F1-score \\
    
    \cite{rs15} & 2019 & DCASE 2016 Dataset \cite{r15} & MFCC & AE, VAE, and VAEGAN & AUC, TPR, and pAUC\\
    
    \cite{rs16} & 2018 & General Sound Effects Library \cite{r14}  & Log mel filter bank & WaveNet, AE, BLSTM-AE, AR-LSTM  & F1-score\\
    
    \cite{rs21} & 2018 & A3FALL \cite{r16} & Log mel-energies, and DWT  & Siamese NN, SVM, One-Class SVM & F1-score, Recall, Precision  \\

    \cite{rs25} & 2018 & TUT Dataset \cite{r13} & MFCC & Elliptic Envelope, and Isolation Forest & F1-score, and ER\\
    
    \cite{rs17} & 2017 &  Own Dataset & Mel-spectrogram & LSTM-AE & ROC \\
    
    \cite{rs18} & 2017 &  Own Dataset, UrbanSound8K \cite{r12} & Mel-spectrogram, Gammatone filterbanks & KNN & ROC, and AUC\\ 

    \cite{rs19} & 2016 & Mivia Dataset \cite{r8} & MFCC, ZCR, Energy ratios in Bark sub-bands, Audio spectrum centroid, Audio spectrum roll-off, Audio spectrum spread & SVM & RR, MDR, ER, FPR, ROC, AUC, Sensitivity \\

    \cite{rs20} & 2014 & Own Dataset & ZCR, FFT, DWT, MFCC, & One-Class SVM &  F1-score, SD, LPC, and LPCC\\

\bottomrule
\label{tab:papersRS2}
\end{tabular}
\end{table}

\begin{table}[htpb]
\caption{Anomaly Detection in audio presented in the included studies of the SR (Manual Search).}
\begin{tabular}{p{0.05\textwidth}p{0.05\textwidth}p{0.15\textwidth}p{0.2\textwidth}p{0.2\textwidth}p{0.15\textwidth}}
\toprule
    {\bf Study} & {\bf Year} & {\bf Dataset} & {\bf Audio Features} & {\bf ML Model} & {\bf Evaluation Method} \\ 
\midrule
    \cite{rs26} & 2020 & ToyADMOS \cite{r9}, MIMII \cite{r10}  & Log-mel energies & ResNet & AUC\\
    
   \cite{rs27} & 2020 & ToyADMOS \cite{r9}, MIMII \cite{r10}  & Gammatone Spectrogram  & AE & ROC, AUC, pAUC\\
    
   \cite{rs28} & 2020 & ToyADMOS \cite{r9}, MIMII \cite{r10}  & Spectrogram & AE & AUC, pAUC \\

  \cite{rs29} & 2020 & ToyADMOS \cite{r9}, MIMII \cite{r10}  &  Log-mel energies & CNN & AUC,  pAUC\\

     \cite{rs30} & 2020 & ToyADMOS \cite{r9}, MIMII \cite{r10}  &  Log-mel energies & CAE & ROC, AUC,  pAUC \\
     
    \cite{rs31} & 2020 & ToyADMOS \cite{r9}, MIMII \cite{r10}   &  Log-mel energies & ResNet, MobileNetV2, GroupMADE  & AUC,  pAUC \\

      \cite{rs32} & 2020 & ToyADMOS \cite{r9}, MIMII \cite{r10}  &  Log-mel energies & CNN, PCA, RLDA, PLDA  & AUC\\
			
\bottomrule
\label{tab:papersRS3}
\end{tabular}
\end{table}

About datasets, 9 studies created and used their own dataset, and  22 studies used public and private datasets. In total, 14 datasets and the main ones were identified, where the 3 most cited datasets were: \textbf{ToyADMOS} \cite{r9}, \textbf{MIMII} \cite{r10}, and \textbf{Mivia} \cite{r8}. The ToyADMOS dataset is a machine operating sound dataset that has approximately 540 hours of normal sound and approximately 12,000 hours of anomalous sound. ToyADMOS was designed to detect audio anomalies in research involving machine operation \cite{r9}. The MIMII dataset is a data set for investigation and inspection of defective industrial machines. It contains the sounds generated from four types of industrial machines (valves, pumps, fans and slide rails) \cite{r10}.Mivia dataset is an audio dataset composed of 6,000 events considered to be vigilance (glass break, shots and screams) \cite{r8}.

In ML, features are the independent variables that serve as input to your ML system or model. The ML model uses features to learn and make predictions. About the audio features method, 34 methods were identified. The main methods of extracting features from the analyzed audio were: \textbf{Mel-frequency cepstral coefficients} (MFCCs), \textbf{Log-Mel Energy}, and \textbf{Mel-spectrogram}.

To answer the main question of this SR, 33 machine learning techniques were identified to detect anomalies in audio.However, two machine learning techniques stand out: the \textbf{Autoencoder} (AE) and the \textbf{Convolutional Neural Network} (CNN). In the most recent studies, the transfer learning method is being used. Trasnfer Learning is an ML method in which a model developed in one task is reused as a starting point in another task. The developed models identified in this study: \textbf{DenseNet-121}, \textbf{MobileNetV2}, and \textbf{ResNet-50}.

About the evaluation method, 23 (approximately 75\%) studies used \textbf{AUC-ROC} and \textbf{F1-score}. AUC-ROC is a performance evaluation that involves classification problems with thresholds. AUC represents a degree of separability and ROC is a probability curve. The higher the AUC, the better the model for predicting a particular class. F1-score measures the accuracy of an ML model. It is widely used in classification, in our example, "normal" and "abnormal".

\newpage

\section{Conclusions}
This paper presents a systematic review on anomaly detection in audio using machine learning techniques. This research had as main objective to obtain the state of the art, enabling an organization of ideas and summarization of information.

In total, 31 studies were selected to study machine learning techniques for anomaly sound detection. After the analysis, 33 machine learning techniques were identified, where AE and CNN were the most cited. We also analyzed the most-used datasets for anomaly detection as their respective methods for extracting features and the evaluation method for machine learning models.

It aims that this study, the result of a secondary study, may allow some direction in the works and research related to the theme. In particular, the author's interest is related to the use of machine learning techniques for anomaly sound detection in-vehicle.

\clearpage
\newpage 

\appendix

\section{Image - Number of authors per year}
\begin{figure}[!ht]
	\centering
	\includegraphics[width=1\linewidth]{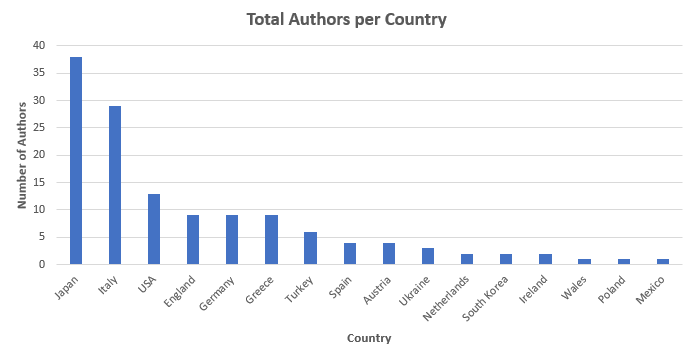}
	\caption{Number of primary of authors per year.}
	\label{fig:allautors}
\end{figure}

\section{Table \ref{tab:allauthor1} - Reseachers of studies}
\begin{table}[htbp]	
	\centering
	\caption{\small{Researchers (author and co-author) and their publications contained in this SR.}}
	{\small
		\begin{tabulary}{\columnwidth}{lclc}	
			\hline 	
			{\bf Author} & {\bf Pub} & {\bf Affiliation} & {\bf Country}   \\ 
			\hline
			Koizumi, Y.	& 4 &	NTT Media Intelligence Laboratories	& Japan \\

			Saito, S. & 3 &	NTT Media Intelligence Laboratories & Japan \\
			
			Uematsu, H. & 3 & NTT Media Intelligence Laboratories & Japan \\
			
			Harada, N. & 3 & NTT Media Intelligence Laboratories & Japan \\
			
			Vafeiadis, A. &	2 &	Center for Research and Technology Hellas-Information  & Greece \\
			& & Technologies Institute \\
			
			Votis, K. &	2 & Center for Research and Technology Hellas-Information  & Greece \\ 
			& & Technologies Institute	\\

			Tzovaras, D. & 2 &	Center for Research and Technology Hellas-Information & Greece \\
			& & Technologies Institute	\\

			Saggese, A.	& 2 & Department of Information Engineering, Electrical & Italy \\
			& & Engineering and Applied Mathematics, University of Salerno \\

            Vento, M. & 2 &	Department of Information Engineering, Electrical & Italy \\
            & & Engineering and Applied Mathematics, University of Salerno \\
            
            Vecchio, M.	& 2 & OpenIoT research unit, FBK CREATE-NET	 & Italy \\
            
             Antonini, M. & 2 &	OpenIoT research unit, FBK CREATE-NET &	Italy \\
             
            Antonelli, F. & 2 &	OpenIoT research unit, FBK CREATE-NET &	Italy \\
            
            Murata, S.	& 2 &	NTT Media Intelligence Laboratories	& Japan \\ 
            
            Kawaguchi, Y. & 2 &	Research and Development Group, Hitachi	& Japan \\ 
            
            Endo, T. & 2 &	Research and Development Group, Hitachi	& Japan \\
            
            Komatsu, T. &	2 &	Data Science Research Laboratories, NEC Corporation	& Japan \\
            
            Hayashi, T.	&2 &	Department of Information Science, Nagoya University	&Japan \\
            
            Kondo, R.	&2&	Data Science Research Laboratories, NEC Corporation	&Japan \\
 
            Toda, T. & 2&	Information Technology Center, Nagoya University &	Japan \\
 
            Takeda, K.	&2&	Department of Information Science, Nagoya University &	Japan \\
            
            Petkov, N.&	2&	Faculty of Science and Engineering, University of Groningen &	Netherlands \\
            
            Bayram, B.	&2&	Faculty of Computer and Informatics Engineering, Istanbul &	Turkey \\
            & & Technical University \\
            
            Duman, T.B.	&2&	Faculty of Computer and Informatics Engineering, Istanbul &	Turkey \\
            & & Technical University \\

            Ince, G.	&2&	Faculty of Computer and Informatics Engineering, Istanbul &	Turkey \\
            & & Technical University \\
            
            \hline
            
            \end{tabulary} 
	}	

	\label{tab:allauthor1}
\end{table}

\begin{table}[htbp]	
	\centering
	\caption{\small{ Continuation of Table 4 (part 1) }}
	{\small
		\begin{tabulary}{\columnwidth}{lclc}	
			\hline 	
			{\bf Author} & {\bf Pub} & {\bf Affiliation} & {\bf Country}   \\ 
			\hline
            
            Plumbley, M. D. &1&	Centre for Vision, Speech and Signal Processing University & England \\
            & &  of Surrey \\
            
            Becker, P.	&1&	FZI Research Center for Information Technology	& Germany \\
            
            Roth, C.	&1&	FZI Research Center for Information Technology	& Germany \\ 
            
            Roennau, A.	&1&	FZI Research Center for Information Technology &	Germany \\
            
            Dillmann, R.	&1&	FZI Research Center for Information Technology	& Germany \\

            Henze, D.	&1&	Technische Universität München	& Germany \\
            
            Gorishti, K. &	1&	Technische Universität München	& Germany \\
            
            Bruegge, B.	&1&	Technische Universität München	& Germany \\
            
            Simen, J.-P.&	1&	Carl Zeiss AG	& Germany\\
            
            Stemmer, G.	&1&	Intel Corp, Intel Labs	& Germany\\
            
            Rushe, E.	&1&	Insight Centre for Data Analytics, University College Dublin	& Ireland\\
            
            Namee, B.M.	&1&	Insight Centre for Data Analytics, University College Dublin	& Ireland \\
            
            Rovetta, S.	&1&	DIBRIS, Universita degli studi di Genova	& Italy\\
            
            Mnasri, Z.	& 1& 	DIBRIS, Universita degli studi di Genova	&Italy\\
            
            Masulli, F. &1&	DIBRIS, Universita degli studi di Genova	& Italy \\
            
            Greco, A.	&1&	Department of Information Engineering, Electrical & Italy \\
            & & Engineering and Applied Mathematics, University of Salerno \\	
            
            Janjua, Z.H.	&1&	OpenIoT research unit, FBK CREATE-NET &	Italy \\

            Foggia, P. 	&1&	Department of Information Engineering, Electrical &  Italy\\ 
            & & Engineering and Applied Mathematics,University of Salerno \\ 
            
            Strisciuglio, N.	&1&	Department of Information Engineering, Electrical &  Italy\\ 
            & & Engineering and Applied Mathematics,University of Salerno \\
            
            Aurino, F.	&1&	Dipartimento di Ingegneria Elettrica e delle Tecnologie & Italy\\ 
            & & dell’Informazione, University of Naples Federico II	\\
            
            Folla, M.	&1&	Dipartimento di Ingegneria Elettrica e delle Tecnologie & Italy\\ 
            & & dell’Informazione, University of Naples Federico II	\\
            
            Gargiulo, F.	&1&	Dipartimento di Ingegneria Elettrica e delle Tecnologie & Italy\\ 
            & & dell’Informazione, University of Naples Federico II	\\
            
            Moscato, V.	&1&	Dipartimento di Ingegneria Elettrica e delle Tecnologie & Italy\\ 
            & & dell’Informazione, University of Naples Federico II	\\
            
            Picariello, A.	& 1&	Dipartimento di Ingegneria Elettrica e delle Tecnologie & Italy\\ 
            & & dell’Informazione, University of Naples Federico II	\\
            
            Sansone, C.	&1&	Dipartimento di Ingegneria Elettrica e delle Tecnologie & Italy\\ 
            & & dell’Informazione, University of Naples Federico II	\\
            
            Droghini, D.	& 1& 	Università Politecnica delle Marche	& Italy \\
            
            Vesperini, F.	&1&	Università Politecnica delle Marche	& Italy \\
            
            Principi, E.	& 1& 	Università Politecnica delle Marche	& Italy \\
            
            Squartini, S.	&1&	Università Politecnica delle Marche	& Italy \\
            
            Piazza, F. &1&	Università Politecnica delle Marche	& Italy\\
            
            Ducange, P.	&1&	SMARTEST Research Centre, eCampus University	& Italy \\
            
            Yasuda, M.	&1&	NTT Media Intelligence Laboratories	& Japan \\
            
            Yamaguchi, M.	&1&	NTT Media Intelligence Laboratories	& Japan \\
            
            Tanabe, R.	&1&	Research and Development Group, Hitachi	& Japan \\
            
            \hline
            \end{tabulary} 
	}	

	\label{tab:allauthorpart1}
\end{table}

\begin{table}[htbp]	
	\centering
	\caption{\small{ Continuation of Table 4 (part 2)  }}
	{\small
		\begin{tabulary}{\columnwidth}{lclc}	
			\hline 	
			{\bf Author} & {\bf Pub} & {\bf Affiliation} & {\bf Country}   \\ 
			\hline
            Ichige, K.	& 1&	Research and Development Group, Hitachi	& Japan \\
            
            Hamada, K.	&1&	Research and Development Group, Hitachi	& Japan \\
            
            Inoue, T.	&1&	IBM Research	& Japan \\
            
            Vinayavekhin, P.	&1&	IBM Research	& Japan\\
            
            Morikuni, S.	&1&	IBM Research	& Japan \\

            Tachibana, R.	&1&	IBM Research	& Japan \\
            
            Lopez-Meyer, P.	&1&	Intel Corp, Intel Labs	& Mexico\\
            
            Kapka, S.	&1&	Samsung R and D Institute Poland & Poland \\
            
            Park, J.	&1&	Advanced Robot Research Laboratory, LG Electronics	& South Korea\\
            
            Yoo, S.	&1&	Advanced Robot Research Laboratory, LG Electronics	& South Korea \\
            
            Perez-Castanos, 	&1&	Visualfy, Benisano	& Spain\\
            
            Naranjo-Alcazar, J.	&1&	Visualfy, Benisano	& Spain \\
            
            Zuccarello, P.	&1&	Visualfy, Benisano	& Spain \\
            
            Cobos, M.	&1&	Universitat de Valencia	& Spain \\
            
            Provotar, O. I.	&1&	Faculty of Computer Science and Cybernetics Taras 	& Ukraine\\
            & & Shevchenko National University of Kyiv \\
            
            Linder, Y. M.	&1&	Faculty of Computer Science and Cybernetics Taras 	& Ukraine\\
            & & Shevchenko National University of Kyiv \\
            
            Veres, M. M. &1&	Faculty of Computer Science and Cybernetics Taras 	& Ukraine\\
            & & Shevchenko National University of Kyiv \\
            
            Giri, R.	&1&	Amazon Web Services	& USA\\
            
            Tenneti, S. V.	&1&	Amazon Web Services	& USA\\
            
            Cheng, F.	&1&	Amazon Web Services	& USA\\
            
            Helwani, K.	&1&	Amazon Web Services	&USA\\
            
            Isik, U.	&1&	Amazon Web Services	& USA\\
            
            Krishnaswamy, A.	&1&	Amazon Web Services	& USA\\
            
            Wang, S.	&1&	IBM Research	&USA \\
            
            Trong, T. H.	&1&	IBM Research	&USA\\
            
            Wood, D.	&1&	IBM Research	&USA\\
            
            Lopez, J. A.	&1&	Intel Corp, Intel Labs	& USA\\
            
            Lu, H.	&1&	Intel Corp, Intel Labs	&USA\\
            
            Nachman, L.	&1&	Intel Corp, Intel Labs	& USA\\
            
            Huang, J.	&1&	Intel Corp, Intel Labs	& USA\\
            
            Perera, C.	&1&	School of Computer Science and Informatics, & Wales\\
            & & Cardiff University	\\
            
             Primus, P.	&1&	Institute of Computational Perception	& Austria \\
            
            Haunschmid, V.	& 1&	Institute of Computational Perception	& Austria\\
            
            Praher, P.	& 1& 	Software Competence Center Hagenberg GmbH	& Austria\\
            
            Widmer, G. &1& 	LIT Artificial Intelligence Lab, Johannes Kepler University &	Austria\\
            
            Chen, L.	&1&	Faculty of Computing, Engineering and Media, De Montfort & England \\ 
            & & University	\\

            Hamzaoui, R.&	1&	Faculty of Computing, Engineering and Media, De Montfort & England \\ 
            & & University	\\
            
            Marchegiani, L.	&1&	Oxford Robotics Institute, University of Oxford	& England \\
            
            Papadimitriou, I.	&1&	Center for Research and Technology Hellas-Information & Greece\\
            & & Technologies Institute	\\
            
            Lalas, A.&	1&	Center for Research and Technology Hellas-Information & Greece\\ 
            & & Technologies Institute	\\
            
            Giakoumis, D. &	1&	Center for Research and Technology Hellas-Information & Greece\\ 
            & & Technologies Institute	\\

			\hline

		\end{tabulary} 
	}	

	\label{tab:allauthorpart2}
\end{table}

\begin{table}[htbp]	
	\centering
	\caption{\small{Continuation of Table 4 (part 3) }}
	{\small
		\begin{tabulary}{\columnwidth}{lclc}	
			\hline 	
			{\bf Author} & {\bf Pub} & {\bf Affiliation} & {\bf Country}   \\ 
			\hline
			
			 Posner, I.	&1&	Oxford Robotics Institute, University of Oxford	& England \\
            
            Kong, Q. &	1&	Centre for Vision, Speech and Signal Processing University & England \\
            & &  of Surrey \\
            
            Xu, Y.	&1&	Centre for Vision, Speech and Signal Processing University & England \\
            & &  of Surrey \\
            
            Sobieraj, I.	&1&	Centre for Vision, Speech and Signal Processing University & England \\
            & &  of Surrey \\
            
            Wang, W.	&1&	Centre for Vision, Speech and Signal Processing University & England \\
            & &  of Surrey \\
            
            \hline

		\end{tabulary} 
	}	

	\label{tab:allauthorpart3}
\end{table}


\newpage

\begin{thebibliography}{99}


\bibitem{r1}
Kawaguchi, Y., \& Endo, T. (2017, September). How can we detect anomalies from subsampled audio signals. In 2017 IEEE 27th International Workshop on Machine Learning for Signal Processing (MLSP) (pp. 1-6). IEEE.

\bibitem{r2}
Henze, D., Gorishti, K., Bruegge, B., \& Simen, J. P. (2019, December). AudioForesight: A Process Model for Audio Predictive Maintenance in Industrial Environments. In 2019 18th IEEE International Conference On Machine Learning And Applications (ICMLA) (pp. 352-357). IEEE.

\bibitem{r3}
Foggia, P., Petkov, N., Saggese, A., Strisciuglio, N., \& Vento, M. (2015). Audio surveillance of roads: A system for detecting anomalous sounds. IEEE transactions on intelligent transportation systems, 17(1), 279-288.

\bibitem{r4}
Alla, S., \& Adari, S. K. (2019). Beginning Anomaly Detection Using Python-Based Deep Learning. Apress.

\bibitem{r5}
 Almeida Biolchini, J. C., Mian, P. G., Natali, A. C. C., Conte, T. U., \& Travassos, G. H. (2007). Scientific research ontology to support systematic review in software engineering. Advanced Engineering Informatics, 21(2), 133-151.

\bibitem{r6}
Brereton, P., Kitchenham, B. A., Budgen, D., Turner, M., \& Khalil, M. (2007). Lessons from applying the systematic literature review process within the software engineering domain. Journal of systems and software, 80(4), 571-583.´

\bibitem{r7} 
dos Santos, A. C. C., Delamaro, M. E., \& Nunes, F. L. (2013, May). The relationship between requirements engineering and virtual reality systems: A systematic literature review. In 2013 XV Symposium on Virtual and Augmented Reality (pp. 53-62). IEEE.

\bibitem{r8}
Foggia, P., Petkov, N., Saggese, A., Strisciuglio, N., \& Vento, M. (2015). Reliable detection of audio events in highly noisy environments. Pattern Recognition Letters, 65, 22-28.

\bibitem{r9}
Koizumi, Y., Saito, S., Uematsu, H., Harada, N., \& Imoto, K. (2019, October). ToyADMOS: A dataset of miniature-machine operating sounds for anomalous sound detection. In 2019 IEEE Workshop on Applications of Signal Processing to Audio and Acoustics (WASPAA) (pp. 313-317). IEEE.

\bibitem{r10}
Purohit, H., Tanabe, R., Ichige, K., Endo, T., Nikaido, Y., Suefusa, K., \& Kawaguchi, Y. (2019). MIMII dataset: Sound dataset for malfunctioning industrial machine investigation and inspection. arXiv preprint arXiv:1909.09347.

\bibitem{r9}
Sammarco, M., \& Detyniecki, M. (2018). Crashzam: Sound-based Car Crash Detection. VEHITS.

\bibitem{r10}
R. (2008, September 21). Pack: KITCHEN common sounds. Freesound.Org. https://freesound.org/people/Robinhood76/packs/3870/

\bibitem{r11}
Yuma Koizumi, Shoichiro Saito, Masataka Yamaguchi, Shin Murata and Noboru Harada, "Batch Uniformization for Minimizing Maximum Anomaly Score of DNN-based Anomaly Detection in Sounds," in Proc of Workshop on Applications of Signal Processing to Audio and Acoustics (WASPAA), 2019.

\bibitem{r12}
Salamon, J., Jacoby, C., \& Bello, J. P. (2014, November). A dataset and taxonomy for urban sound research. In Proceedings of the 22nd ACM international conference on Multimedia (pp. 1041-1044).

\bibitem{r13}
Aleksandr Diment, Annamaria Mesaros, Toni Heittola, \& Tuomas Virtanen. (2018). TUT Rare sound events, Evaluation dataset [Data set]. Zenodo. http://doi.org/10.5281/zenodo.1160455

\bibitem{r14}
 “Series 6000 general sound effects library,” http://www.sound-ideas.com/sound-effects/series-6000-sound-effects-library.html,
[Accessed: 02- Jan- 2021]

\bibitem{r15}
DCASE2016 Challenge - DCASE. (2016).  http://dcase.community/challenge2016/index

\bibitem{r16}
Principi, E., Droghini, D., Squartini, S., Olivetti, P.,  Piazza, F. (2016). Acoustic cues from the floor: a new approach for fall classification. Expert Systems with Applications, 60, 51-61.

\bibitem{r17}
A. Mesaros, T. Heittola, and T. Virtanen, “A multi-device dataset for urban
acoustic scene classification,” in Proc. Detect. Classif. Acoust. Sce. Events
2018 Workshop, 2018, pp. 9–13

\bibitem{r18}
E. Fonseca et al., “General-purpose tagging of freesound audio
with audioset labels: Task description, dataset, and baseline,” 2018,arXiv:1807.09902

\bibitem{r19}
Various artificial signal datasets, https://github.com/numenta/NAB/tree/master/data

\bibitem{r20}
Fabbri, S., Hernandes, E., Di Thommazo, A., Belgamo, A., Zamboni, A., \& Silva, C. (2012, June). Using information visualization and text mining to facilitate the conduction of systematic literature reviews. In International Conference on Enterprise Information Systems (pp. 243-256). Springer, Berlin, Heidelberg.


 
 \bibitem{rs1}
 Papadimitriou, I., Vafeiadis, A., Lalas, A., Votis, K., \& Tzovaras, D. (2020). Audio-Based Event Detection at Different SNR Settings Using Two-Dimensional Spectrogram Magnitude Representations. Electronics, 9(10), 1593.

\bibitem{rs2}
Koizumi, Y., Yasuda, M., Murata, S., Saito, S., Uematsu, H., \& Harada, N. (2020, May). SPIDERnet: Attention Network For One-Shot Anomaly Detection In Sounds. In ICASSP 2020-2020 IEEE International Conference on Acoustics, Speech and Signal Processing (ICASSP) (pp. 281-285). IEEE.

\bibitem{rs3}
Rovetta, S., Mnasri, Z., \& Masulli, F. (2020, May). Detection of hazardous road events from audio streams: An ensemble outlier detection approach. In 2020 IEEE Conference on Evolving and Adaptive Intelligent Systems (EAIS) (pp. 1-6). IEEE.

\bibitem{rs4}
Becker, P., Roth, C., Roennau, A., \& Dillmann, R. (2020, April). Acoustic Anomaly Detection in Additive Manufacturing with Long Short-Term Memory Neural Networks. In 2020 IEEE 7th International Conference on Industrial Engineering and Applications (ICIEA) (pp. 921-926). IEEE.

\bibitem{rs5}
Vafeiadis, A., Votis, K., Giakoumis, D., Tzovaras, D., Chen, L., \& Hamzaoui, R. (2020). Audio content analysis for unobtrusive event detection in smart homes. Engineering Applications of Artificial Intelligence, 89, 103226.

\bibitem{rs6}
Bayram, B., Duman, T. B., \& Ince, G. (2020). Real time detection of acoustic anomalies in industrial processes using sequential autoencoders. Expert Systems, e12564.

\bibitem{rs7}
Duman, T. B., Bayram, B., \& İnce, G. (2019, May). Acoustic Anomaly Detection Using Convolutional Autoencoders in Industrial Processes. In International Workshop on Soft Computing Models in Industrial and Environmental Applications (pp. 432-442). Springer, Cham.

\bibitem{rs8}
Greco, A., Petkov, N., Saggese, A., \& Vento, M. (2020). AReN: A Deep Learning Approach for Sound Event Recognition using a Brain inspired Representation. IEEE Transactions on Information Forensics and Security.

\bibitem{rs9}
Henze, D., Gorishti, K., Bruegge, B., \& Simen, J. P. (2019, December). AudioForesight: A Process Model for Audio Predictive Maintenance in Industrial Environments. In 2019 18th IEEE International Conference On Machine Learning And Applications (ICMLA) (pp. 352-357). IEEE.

\bibitem{rs10}
Koizumi, Y., Saito, S., Yamaguchi, M., Murata, S., \& Harada, N. (2019, October). Batch uniformization for minimizing maximum anomaly score of dnn-based anomaly detection in sounds. In 2019 IEEE Workshop on Applications of Signal Processing to Audio and Acoustics (WASPAA) (pp. 6-10). IEEE.

\bibitem{rs11}
Janjua, Z. H., Vecchio, M., Antonini, M., \& Antonelli, F. (2019). IRESE: An intelligent rare-event detection system using unsupervised learning on the IoT edge. Engineering Applications of Artificial Intelligence, 84, 41-50.

\bibitem{rs12}
Rushe, E., \& Mac Namee, B. (2019, May). Anomaly detection in raw audio using deep autoregressive networks. In ICASSP 2019-2019 IEEE International Conference on Acoustics, Speech and Signal Processing (ICASSP) (pp. 3597-3601). IEEE.

\bibitem{rs13}
Kawaguchi, Y., Tanabe, R., Endo, T., Ichige, K., \& Hamada, K. (2019, May). Anomaly detection based on an ensemble of dereverberation and anomalous sound extraction. In ICASSP 2019-2019 IEEE International Conference on Acoustics, Speech and Signal Processing (ICASSP) (pp. 865-869). IEEE.

\bibitem{rs14}
Komatsu, T., Hayashiy, T., Kondo, R., Todaz, T., \& Takeday, K. (2019, May). Scene-dependent Anomalous Acoustic-event Detection Based on Conditional Wavenet and I-vector. In ICASSP 2019-2019 IEEE International Conference on Acoustics, Speech and Signal Processing (ICASSP) (pp. 870-874). IEEE.

\bibitem{rs15}
Koizumi, Y., Saito, S., Uematsu, H., Kawachi, Y., \& Harada, N. (2018). Unsupervised detection of anomalous sound based on deep learning and the Neyman–Pearson lemma. IEEE/ACM Transactions on Audio, Speech, and Language Processing, 27(1), 212-224

\bibitem{rs16}
Hayashi, T., Komatsu, T., Kondo, R., Toda, T., \& Takeda, K. (2018, September). Anomalous sound event detection based on wavenet. In 2018 26th European Signal Processing Conference (EUSIPCO) (pp. 2494-2498). IEEE.

\bibitem{rs17}
Kawaguchi, Y., \& Endo, T. (2017, September). How can we detect anomalies from subsampled audio signals?. In 2017 IEEE 27th International Workshop on Machine Learning for Signal Processing (MLSP) (pp. 1-6). IEEE.

\bibitem{rs18}
Marchegiani, L., \& Posner, I. (2017, May). Leveraging the urban soundscape: Auditory perception for smart vehicles. In 2017 IEEE International Conference on Robotics and Automation (ICRA) (pp. 6547-6554). IEEE.

\bibitem{rs19}
Foggia, P., Petkov, N., Saggese, A., Strisciuglio, N., \& Vento, M. (2015). Audio surveillance of roads: A system for detecting anomalous sounds. IEEE transactions on intelligent transportation systems, 17(1), 279-288.

\bibitem{rs20}
Aurino, F., Folla, M., Gargiulo, F., Moscato, V., Picariello, A., \& Sansone, C. (2014, September). One-class SVM based approach for detecting anomalous audio events. In 2014 International Conference on Intelligent Networking and Collaborative Systems (pp. 145-151). IEEE.

\bibitem{rs21}
Droghini, D., Vesperini, F., Principi, E., Squartini, S., \& Piazza, F. (2018, August). Few-shot siamese neural networks employing audio features for human-fall detection. In Proceedings of the International Conference on Pattern Recognition and Artificial Intelligence (pp. 63-69).

\bibitem{rs22}
Kong, Q., Xu, Y., Sobieraj, I., Wang, W., \& Plumbley, M. D. (2019). Sound event detection and time–frequency segmentation from weakly labelled data. IEEE/ACM Transactions on Audio, Speech, and Language Processing, 27(4), 777-787.

\bibitem{rs24}
Provotar, O. I., Linder, Y. M., \& Veres, M. M. (2019, December). Unsupervised Anomaly Detection in Time Series Using LSTM-Based Autoencoders. In 2019 IEEE International Conference on Advanced Trends in Information Theory (ATIT) (pp. 513-517). IEEE.

\bibitem{rs25}
Antonini, M., Vecchio, M., Antonelli, F., Ducange, P., \& Perera, C. (2018). Smart audio sensors in the internet of things edge for anomaly detection. IEEE Access, 6, 67594-67610.

\bibitem{rs26}
Primus, P., Haunschmid, V., Praher, P., \& Widmer, G. (2020). Anomalous Sound Detection as a Simple Binary Classification Problem with Careful Selection of Proxy Outlier Examples. arXiv preprint arXiv:2011.02949.

\bibitem{rs27}
Perez-Castanos, S., Naranjo-Alcazar, J., Zuccarello, P., \& Cobos, M. (2020). Anomalous Sound Detection using unsupervised and semi-supervised autoencoders and gammatone audio representation. arXiv preprint arXiv:2006.15321.

\bibitem{rs28}
Park, J., \& Yoo, S. DCASE 2020 TASK2: ANOMALOUS SOUND DETECTION USING RELEVANT SPECTRAL FEATURE AND FOCUSING TECHNIQUES IN THE UNSUPERVISED LEARNING SCENARIO.

\bibitem{rs29}
Inoue, T., Vinayavekhin, P., Morikuni, S., Wang, S., Trong, T. H., Wood, D., ... \& Tachibana, R. (2020). Detection of Anomalous Sounds for Machine Condition Monitoring using Classification Confidence (Vol. 2). Tech. report in DCASE2020 Challenge Task.

\bibitem{rs30}
Kapka, S. (2020). ID-Conditioned Auto-Encoder for Unsupervised Anomaly Detection. arXiv preprint arXiv:2007.05314.

\bibitem{rs31}
Giri, R., Tenneti, S. V., Cheng, F., Helwani, K., Isik, U., \& Krishnaswamy, A. SELF-SUPERVISED CLASSIFICATION FOR DETECTING ANOMALOUS SOUNDS.

\bibitem{rs32}
Wilkinghoff, K. USING LOOK, LISTEN, AND LEARN EMBEDDINGS FOR DETECTING ANOMALOUS SOUNDS IN MACHINE CONDITION MONITORING.



\end{thebibliography}
\end{document}